\documentclass[aps,prb,twocolumn,floatfix,longbibliography]{revtex4-2}
\usepackage{color}
\usepackage{xcolor}
\usepackage{amssymb}
\usepackage{amsmath}
\usepackage{bm}
\usepackage{hyperref}
\usepackage{graphicx}
\usepackage{amsmath,amssymb,bm,epsfig,color}

\makeatother

\begin{document}
\title{Microscopic lattice model for quartic semi-Dirac fermions in two dimensions}
\author{Mohamed M. Elsayed}
\author{Valeri N. Kotov}
\affiliation{Department of Physics, University of Vermont, Burlington, Vermont 05405, USA}
\begin{abstract}
We propose a lattice model for the realization of exotic quartic semi-Dirac fermions, i.e. quasiparticles exhibiting a dispersion with quartic momentum dependence in a given direction, and a linear dependence in the perpendicular direction. A tight binding model is employed, allowing for hopping between up to fourth nearest neighbors and anisotropic hopping parameters. In addition, we introduce short range electron-electron interactions  which are necessary to stabilize the quartic semi-Dirac phase. Without interactions, or in the presence of long range correlations,  the lattice is unstable towards formation of either anisotropic Dirac cones, or simple (quadratic) semi-Dirac phase.
\end{abstract}
\maketitle
\section{Introduction}
Semi-Dirac fermions are strongly anisotropic two dimensional quasiparticles that disperse relativistically in one direction and quadratically along the orthogonal direction. This can be understood as arising from a merger of two Dirac cones at a Lifshitz transition \cite{montambaux2009universal,montambauxmerging}, and can manifest in a variety of physical systems. 
The application of strain to honeycomb lattices is a simple way, in principle, to facilitate such topological transitions \cite{Amorim2016,Wunsch2008}, and is generally a useful tunable parameter that can control interactions between atoms and two-dimensional (2D) materials  \cite{Kim2024,Nichols2016,Elsayed2023}. This is not the only way to bring about such transitions, which are achievable in a variety of Dirac systems with appropriately tunable parameters. While the prototypical platform is a honeycomb lattice, semi-Dirac dispersions are possible in other settings, for example cold atoms on a square lattice can be manipulated to produce a merger of Dirac cones \cite{Delplace2010}. Other potential mechanisms include tuning the interplay between short range interactions and disorder \cite{wang2018role}, or leveraging the response to time-varying electromagnetic fields \cite{Saha2016}.

Semi-Dirac fermions have a density of states $\rho(\varepsilon) \sim\varepsilon^{1/2}$ and  Landau level  magnetic field dependence $\varepsilon_n(B)\sim \pm [(n+1/2)B]^{2/3}$, where $n=0,1,2,...$ \cite{Dietl2008}. The hybrid nature of the spectrum is reflected in the scaling of $\rho(\varepsilon)$, where the power of $\varepsilon$ is always between $0$, as it is for the free 2D electron gas, and $1$, as it is for simple Dirac fermions. 
Interactions have been shown to have strong effects in such systems, and two distinct low energy regimes have been identified in the weak coupling limit. A large $N_f$ (number of fermion species) approach yields particularly strong logarithmic corrections \cite{isobe2016}, while a perturbative renormalization group treatment produces a novel resummation of  $\log^2$ divergences to all orders in the interaction strength that restores the linear dispersion near the band crossing \cite{kotov2021,elsayed2025}. 

Furthermore, there are type-II semi-Dirac fermions, which disperse in the same fashion along the principal axes but exhibit an admixture of different momentum components at intermediate angles. This is a symmetry-reduced model that was suggested to explain the non-trivial topological properties observed in some semi-Dirac systems, and can be understood as resulting from a merger of three Dirac cones, producing a non-zero Berry phase \cite{huang2015}. The density of states and Landau levels for these quasiparticles follows the scaling $\rho(\varepsilon)\sim\varepsilon^{1/3}$ and $\varepsilon_n(B)\sim \pm  (nB)^{3/4}$ respectively.

In addition, it is worth mentioning that exotic cubic, quartic and higher Dirac fermion excitations, with isotropic dispersion (of the type $ \varepsilon{({\bf k})} \sim |{\bf k}|^N$, where $N=3,4,...$), are possible in graphene stacks  \cite{Paco2006,Paco2007,Volovik2012}, leading to peculiar electronic properties. 

In this paper we explore the possibility of realizing an even more exotic flavor of semi-Dirac fermions in which the quadratic term 
in the $k_x$ direction vanishes and the dispersion is quartic to leading order, of the type $|\varepsilon({\bf k})|=\sqrt{(gk_x^4)^2+(vk_y)^2}$.
This spectrum leads to unusual behavior of the physical observables, with the density of states and semiclassical  Landau levels scaling as $\rho(\varepsilon)\sim\varepsilon^{1/4}$ and $\varepsilon_n(B) \sim \pm [(n+1/2)B]^{4/5}$. 
We expect that studies of magneto-optical, polaritonic and superconducting properties performed on conventional semi-Dirac systems  \cite{Zhou2021,Real2020,Uchoa2017}, could be potentially extended to exotic higher order fermions, thus probing their unconventional electronic dispersion and electromagnetic field response. Note that such higher order semi-Dirac models are topologically trivial, in the sense that the Berry phase, and consequently the Chern number, are identically zero.
Higher order semi-Dirac fermions have been abstractly considered in the literature \cite{elsayed2025,roy2018,quan2018maximally,carbotte2019} where various effects and properties have been studied, but hitherto there has been no suggestion of a microscopic model that could potentially host such excitations. We show that this is indeed possible in our proposed model if electron-electron interactions are taken into account. Short range interactions are necessary to stabilize the quartic semi-Dirac phase, whereas long range Coulomb interactions induce an ordinary semi-Dirac dispersion.

In the rest of the paper we first construct a lattice model (Section \ref{sec:model}), with nearest and further neighbor hopping parameters (up to fourth neighbors). Section \ref{sec:interactions}
 contains a discussion of the crucial role short range electron-electron interactions can play in the emergence of a quartic semi-Dirac spectrum. In Section \ref{sec:conclusions} we present our conclusions. Appendix \ref{AppendixA} contains analysis of long range Coulomb interactions.
 
\section{Model}
\label{sec:model}
\begin{figure}[b]
    \centering
    \includegraphics[width=0.9\linewidth]{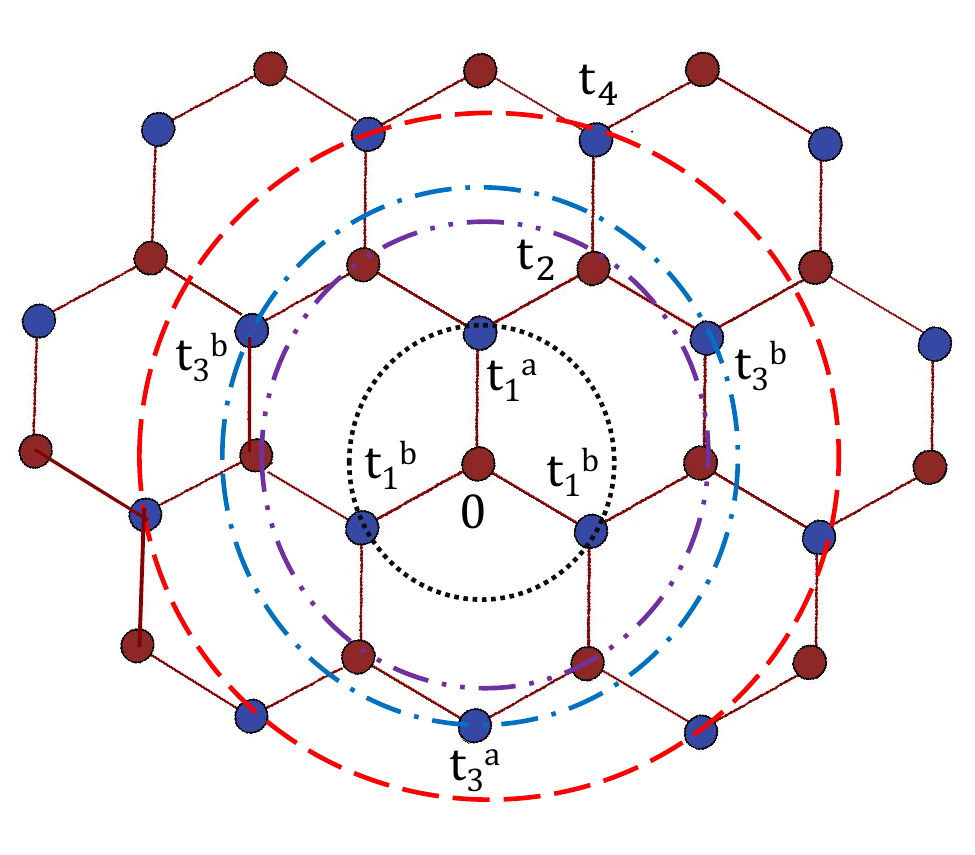}
    \caption{Honeycomb lattice with up to fourth nearest neighbors indicated by the concentric circles. We consider anisotropic first and third nearest neighbor hoppings $t_1^a,t_1^b,t_3^a,t_3^b$, and an isotropic fourth nearest neighbor hopping $t_4.$}
    \label{fig:lattice}
\end{figure}
We consider a tight-binding model on a 2D honeycomb lattice with up to fourth nearest neighbor hoppings. The Bloch hamiltonian reads
\begin{equation}
    H_k=\begin{pmatrix}
        \varepsilon_2(\mathbf{k})&& \varepsilon_1(\mathbf{k})+\varepsilon_3(\mathbf{k})+\varepsilon_4(\mathbf{k})\\
        \varepsilon_1^{\ast}(\mathbf{k})+\varepsilon_3^{\ast}(\mathbf{k})+\varepsilon_4^{\ast}(\mathbf{k})&&\varepsilon_2(\mathbf{k})
    \end{pmatrix}
\end{equation}
where
\begin{equation}
    \varepsilon_i=-\sum_{j} t_{i}^je^{i\mathbf{k}\cdot\mathbf{R}_{j}}
\end{equation}
and the sum runs over vectors $\mathbf{R}_{j}$ connecting the $i$th nearest neighbors, and $t_{i}^j$ is the corresponding hopping integral. The spectrum is easily found to be
\begin{equation}
    \varepsilon(\mathbf{k})=\varepsilon_2(\mathbf{k})\pm\vert\varepsilon_1(\mathbf{k})+\varepsilon_3(\mathbf{k})+\varepsilon_4(\mathbf{k})\vert.
\end{equation}
The diagonal entries corresponding to hopping between the same sublattice can act as an effective doping or to tilt the Dirac cones, but do not affect the forthcoming analysis and will be neglected in what follows. It is known that manipulation of the $t_{i}^j$ can lead to motion and nucleation of Dirac points in momentum space \cite{montambauxmerging,bena2011}. For example, assuming isotropic hoppings ($t_i^j\to t_i$), and only considering up to third nearest-neighbor hopping ($t_4=0$), it can be shown that a pair of Dirac points are created at the midpoints of the edges of the Brillouin zone, usually called $M\text{-points}$, under the condition $t_3=t_1/3$ \cite{bena2011}. Expanding the spectrum around one of these points shows that the velocity vanishes along the direction connecting the $K$ and $K^{\prime}$ points and the leading behavior is quadratic, giving rise to an ordinary semi-Dirac dispersion. Alternatively, a semi-Dirac spectrum can emerge considering only nearest neighbor hopping if the hopping integrals are allowed to be anisotropic $t_1\to t_1^a\neq t_1^b=t_1^c$. In this scenario the existing Dirac points approach each other as $t_1^a$ grows relative to $t_1^b$, and coalesce at at a subset of $M\text{-points}$ when $t_1^a=2t_1^b$ \cite{montambaux2009universal}, producing four semi-Dirac points in the Brillouin zone. If we are to find quartic semi-Dirac fermions, more degrees of freedom are needed, thus we allow for an isotropic, fourth neighbor hopping ($t_4\neq0$) and anisotropic first and third neighbor hoppings $t_1^a\neq t_1^b=t_1^c$, and $t_3^a\neq t_3^b=t_3^c$. 
The lattice and hopping pattern are shown in Figure \ref{fig:lattice}.
Expanding $\gamma(\mathbf{k})\equiv\varepsilon_1+\varepsilon_3+\varepsilon_4$ around $(0,2\pi/3)$, and setting the lattice constant to unity, we find the real and imaginary parts of $\gamma$ to be:
\begin{eqnarray}
    \Re(\gamma)&=&\frac{1}{2}(t_1^a-2t_1^b+t_3^a+2t_3^b-t_4)\nonumber \\& &+\frac{\sqrt{3}}{2}(t_1^a+t_1^b-2t_3^a+2t_3^b-8t_4)k_y\nonumber \\
    & &+\frac{1}{8}(-2t_1^a+t_1^b-8t_3^a-4t_3^b+10t_4)k_y^2 \nonumber \\
    & & +\frac{\sqrt{3}}{48}(-4t_1^a-t_1^b+32t_3^a-8t_3^b+188t_4)k_y^3 \nonumber \\
    & &+\frac{1}{384}(8t_1^a-t_1^b+128t_3^a+16t_3^b-370t_4)k_y^4 \nonumber \\
    & &+\frac{3}{8}(t_1^b-4t_3^b+6t_4)k_x^2 \nonumber \\
    & &+\frac{3\sqrt{3}}{16}(-t_1^b-8t_3^b+12t_4)k_x^2k_y \nonumber \\
    & &+\frac{3}{64}(-t_1^b+16t_3^b+30t_4)k_x^2k_y^2 \nonumber \\
    & &+\frac{3}{128}(-t_1^b+16t_3^b-66t_4)k_x^4 \nonumber
    \end{eqnarray}
    \begin{eqnarray}
    \Im(\gamma)&=&\frac{\sqrt{3}}{2}(-t_1^a+2t_1^b-t_3^a-2t_3^b+2t_4) \nonumber \\
    & &+\frac{1}{2}(t_1^a+t_1^b-2t_3^a+2t_3^b-8t_4)k_y \nonumber \\
    & &+\frac{\sqrt{3}}{8}(2t_1^a-t_1^b+8t_3^a+4t_3^b-10t_4)k_y^2 \nonumber \\
    & &+\frac{1}{48}(-4t_1^a-t_1^b+32t_3^a-8t_3^b+188t_4)k_y^3 \nonumber \\
    & &+\frac{\sqrt{3}}{384}(-8t_1^a+t_1^b-128t_3^a-16t_3^b+370t_4)k_y^4 \nonumber \\
    & &+\frac{3\sqrt{3}}{8}(-t_1^b+4t_3^b-6t_4)k_x^2 \nonumber \\
    & &+\frac{3}{16}(-t_1^b-8t_3^b+12t_4)k_x^2k_y \nonumber \\
    & &+\frac{3\sqrt{3}}{64}(t_1^b-16t_3^b-30t_4)k_x^2k_y^2 \nonumber \\
    & &+\frac{3\sqrt{3}}{128}(t_1^b-16t_3^b+66t_4)k_x^4.
    \label{eq:expansion}
\end{eqnarray}    
We only retain terms up to fourth order in momentum. First we insist that the values of the parameters are such that the terms proportional to $k_y^2,k_y^4,k_x^2,k_x^2k_y^2$ vanish. This leads to the set of conditions:
\begin{equation}
t_1^a=-14t_4,t_1^b=-18t_4,t_3^a=4t_4,t_3^b=-3t_4.
\label{conditions}
\end{equation}
On this curve in parameter space, the hamiltonian becomes:
\begin{eqnarray}
    H&=&\left(9t_4-27\sqrt{3}t_4 k_y + \frac{69\sqrt{3}}{8}t_4 k_y^3+\frac{81\sqrt{3}}{8}t_4 k_x^2 k_y\right. \nonumber \\
    & & \left. -\frac{9}{4}t_4 k_x^4\right)\sigma_x+\left(9\sqrt{3}t_4+27t_4k_y-\frac{69}{8}t_4k_y^3\right. \nonumber \\
    & &\left.-\frac{81}{8}t_4k_x^2k_y-\frac{9\sqrt{3}}{4}t_4k_x^4\right)\sigma_y, 
    \label{eq:ham0}
\end{eqnarray}
where $\sigma_{x,y}$ are the standard Pauli matrices. We may exploit the SU(2) sublattice symmetry of the system to perform a rotation by multiplying the hamiltonian Eq.(\ref{eq:ham0}) by the unitary matrix
\begin{equation}
 U=\begin{pmatrix}
    e^{i\frac{\pi}{3}}&&0\\
    0&& e^{-i\frac{\pi}{3}}
 \end{pmatrix},   
\end{equation}
conveniently shifting all the $k_y$ dependent terms from the $\sigma_x$ to the $\sigma_y$ channel, and removing the constant and $k_x^4$ term from the $\sigma_y$ channel. Given that the leading order in $\sigma_y$ is linear in $k_y$, we may neglect the higher order terms, leaving us with the greatly simplified hamiltonian:
\begin{equation}
    H=\left(-\frac{9}{2}t_4k_x^4+18t_4\right)\sigma_x+(54t_4k_y)\sigma_y.
    \label{ham1}
\end{equation}
\begin{figure}
    \centering
    \includegraphics[width=0.85\linewidth]{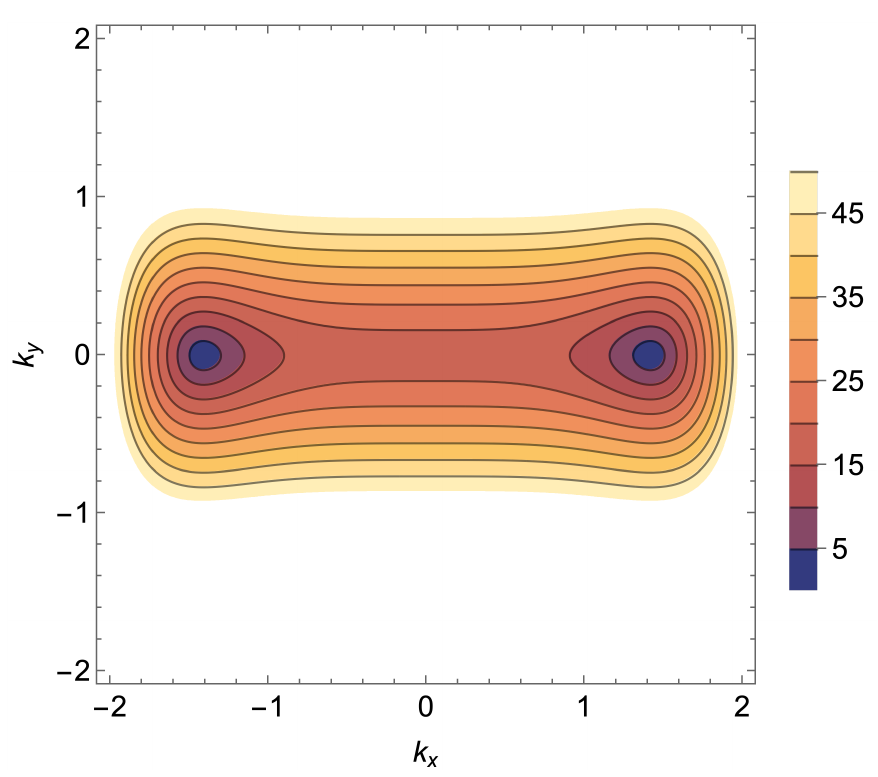}
    \caption{Contour plot of the spectrum corresponding to the hamiltonian in Eq.(\ref{ham1}), where the color bar indicates energy values $\varepsilon(\mathbf{k}).$ The presence of $\Delta$ means the lowest energy excitations are two anisotropic Dirac cones close to $(0,2\pi/3)$, and quartic behavior is only observed at higher energies.  }
    \label{fig:plotwithdelta}
\end{figure}
\begin{figure}
    \centering
    \includegraphics[width=0.9\linewidth]{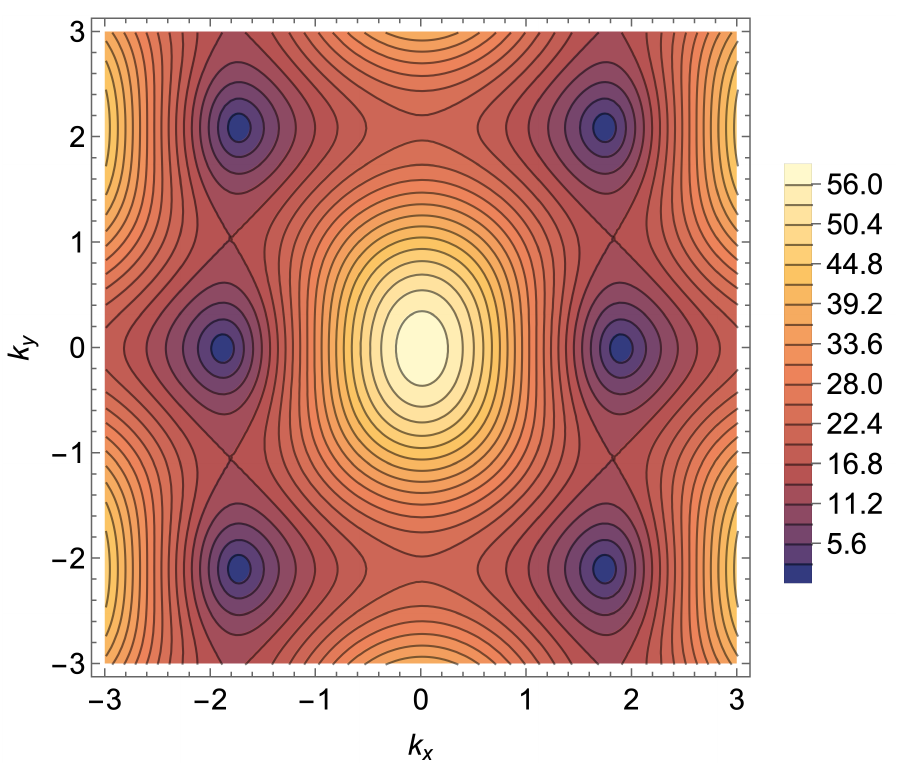}
    \caption{Full band structure under the set of conditions in Eq.(\ref{conditions}) and $t_4=1$. Note the slightly different positions of the Dirac points as compared to Figure \ref{fig:plotwithdelta} (see text)}
    \label{fig:FullBZ}
\end{figure}

This manipulation brings us very close to a purely quartic semi-Dirac dispersion, but not quite, since the constant term $\Delta\equiv18t_4$ survives in $\sigma_x$. The fact that $\Delta$ and $g\equiv -\frac{9}{2}t_4$ have opposite signs indicates that there are two anisotropic Dirac cones nearby, located at $\left(\pm(\Delta/g)^{1/4},2\pi/3\right)$, and the quartic behavior is only present at higher energies as can be seen in Figure \ref{fig:plotwithdelta}. Without making use of a low energy expansion, the full band structure is illustrated in Figure \ref{fig:FullBZ} for the values of the parameters in Eq.(\ref{conditions}). Note that the locations of the Dirac cones are slightly different from those in Figure \ref{fig:plotwithdelta}, since the expansion leading to Eq.(\ref{ham1}) is only accurate for $\vert\mathbf{k}\vert<1$ as measured from $(0,2\pi/3)$, and $(\Delta/g)^{1/4}=\sqrt{2}$.
To negate the presence of $\Delta$ in Eq.(\ref{ham1}) and observe a truly quartic semi-Dirac point, we must turn on interactions.

\section{Electron-electron interactions}
\label{sec:interactions}

\begin{figure}
    \centering
    \includegraphics[width=0.6\linewidth]{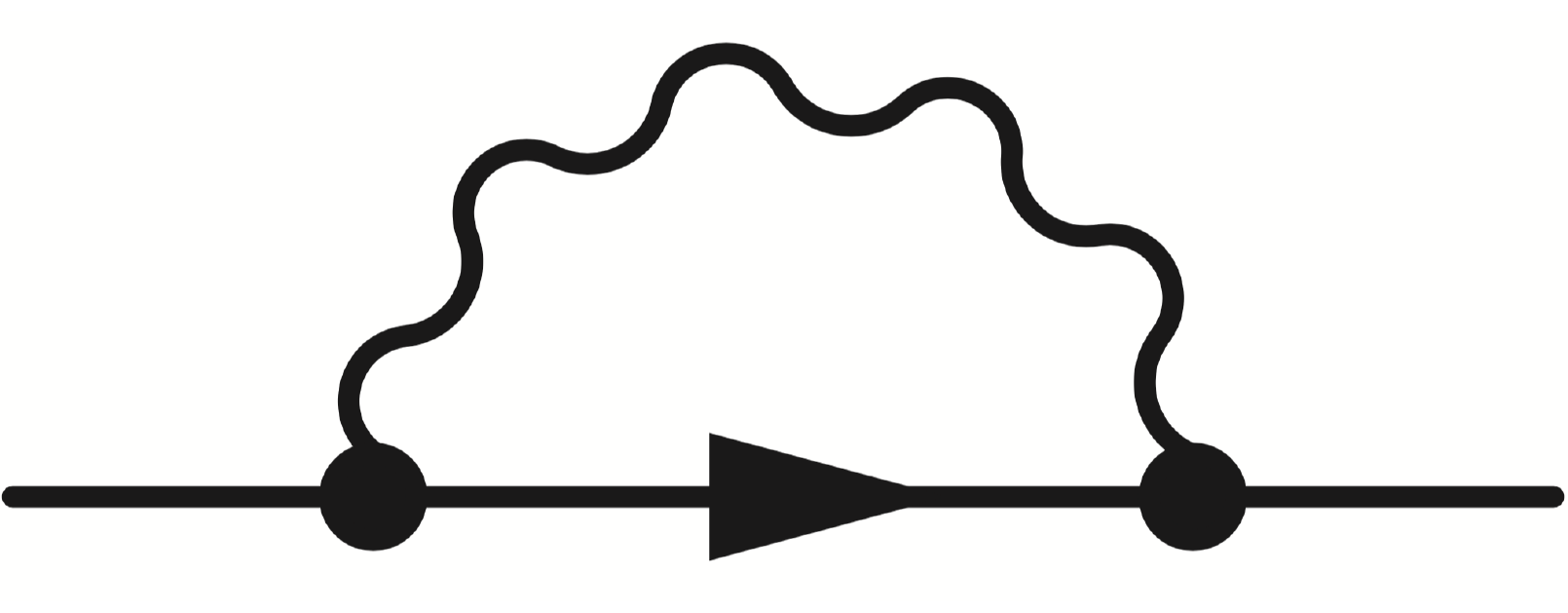}
    \caption{One loop self energy diagram. The wavy line represents the short range interaction $V(\mathbf{k})=U$, or the long range Coulomb potential $V(\mathbf{k})=2\pi e^2/k$.}
    \label{feynman}
\end{figure}

 We consider a short-range electron-electron interaction in the continuum limit $V({\bf r})=U\delta({\bf r})$, which is constant in momentum space $V(\mathbf{k})=U$.
Taking a perturbative weak-coupling approach in the dimensionless interaction strength $u = U/(va) \ll 1$, we calculate the first order (Hartree-Fock) self energy shown in Figure \ref{feynman}
\begin{equation}
    \hat{\Sigma}({\bf p})=\frac{i}{(2\pi)^{3}}\int_{-\infty}^{\infty}\text{d}\omega\iint\text{d}^{2}{\bf k}\,\hat{G}_{0}({\bf k},\omega)V({\bf k}-{\bf p}),
    \label{selfEnergy}
\end{equation}
where 
\begin{equation}
    \hat{G}_{0}({\bf k},\omega)=\left[\omega-H({\bf k})+i0^{+}\text{sgn}(\omega)\right]^{-1}
\end{equation}
is the free Green's function. In Eq. (\ref{selfEnergy}), ${\bf p}$ is the external momentum. The lattice spacing $a$ and Planck's constant are set to one, $\hbar=a=1$.

 The self energy under the contact interaction does not produce a momentum-dependent renormalization, and simply generates the constant:
\begin{equation}
\hat{\Sigma}(\mathbf{p})=\left( \iint\frac{\text{d}^{2}{\bf k}}{(2\pi)^2}\frac{h_{x}(\mathbf{k})}{2\varepsilon({\bf k})}\;U\right)\sigma_x,
\label{eq:SE}
\end{equation}
where  the electronic dispersion is $\varepsilon({\bf k})=\sqrt{h_{x}^{2}(\mathbf{k})+h_{y}^{2}(\mathbf{k})}$, and we define:
\begin{eqnarray}
    h_x(\mathbf{k})&=&-\frac{9}{2}t_4k_x^4+18t_4\equiv gk_x^4+\Delta \nonumber \\
    h_y(\mathbf{k})&=&54t_4k_y\equiv vk_y.
    \label{generalham}
\end{eqnarray}
\begin{figure}
    \centering
    \includegraphics[width=0.9\linewidth]{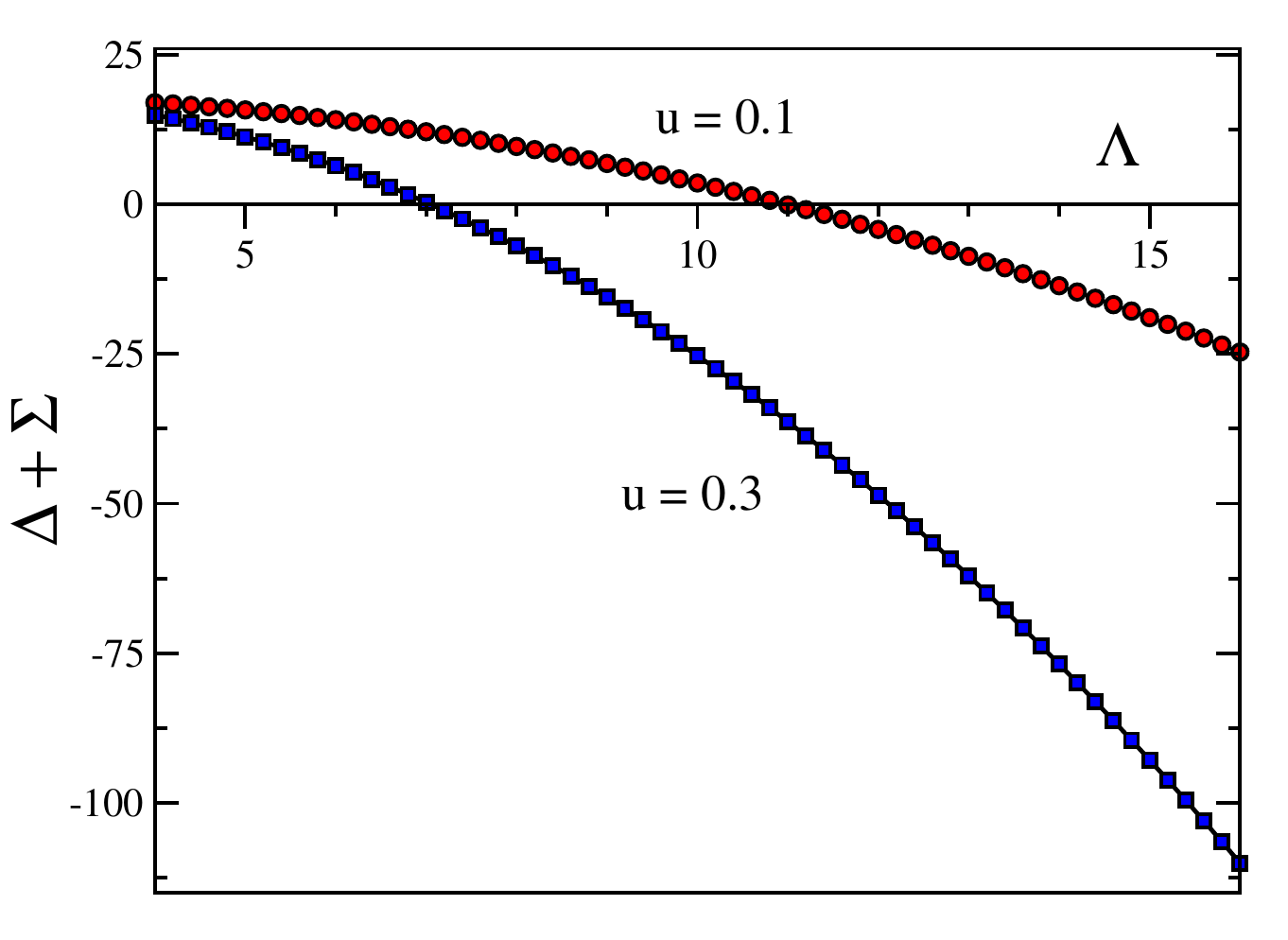}
    \caption{Constant term in the interacting hamiltonian plotted against the ultraviolet momentum
     cutoff $\Lambda$ for different 
    short-range interaction strengths $u=0.1,0.3$. The energies ($y$-axis) are in units of $t_4$ and $\Lambda$ is in units of inverse lattice spacing $a=1$.}
    \label{fig:selfenergyplot2}
\end{figure}
\begin{figure}
    \centering
    \includegraphics[width=0.9
    \linewidth]{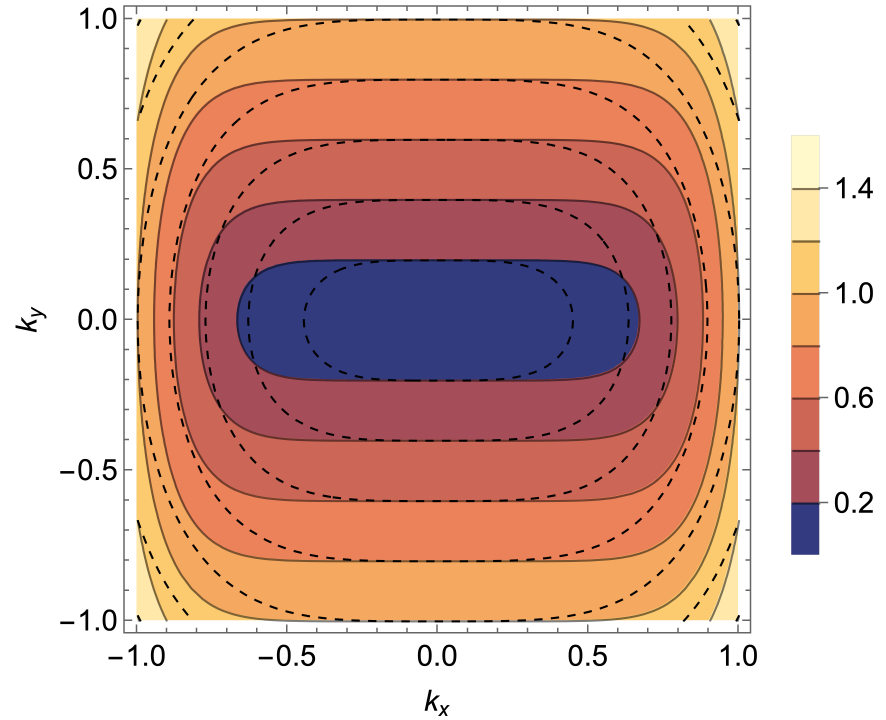}
    \caption{Quartic semi-Dirac dispersion around the point $(0,2\pi/3)$. These exotic low energy excitations can be stabilized by short range interactions driving a topological transition from the Dirac cones of Figure \ref{fig:plotwithdelta} to the novel higher order semi-Dirac spectrum shown above. The superimposed dashed contours reflect the ordinary quadratic semi-Dirac spectrum, included for comparison.}
    \label{fig:quarticspectrum}
\end{figure}
Subsequently $H({\bf p}) \rightarrow H({\bf p}) + \hat{\Sigma}(\mathbf{p})$. 
Only the $\sigma_x$ part of the self-energy is non-zero since $h_y(\mathbf{k})$
is an odd function of momentum. 
Denoting by $\Sigma$ the constant scalar piece of the self-energy (i.e. $\hat{\Sigma}(\mathbf{p})= \Sigma\,\sigma_x$), we observe that the constant  $\Delta=18t_4$  in the hamiltonian  Eq.(\ref{ham1}) can be canceled by $\Sigma$, as shown in Figure \ref{fig:selfenergyplot2}. This is due to the different signs of the two terms in $h_x(\mathbf{k})$. We observe that the effective vanishing of  $\Delta$ in Figure \ref{fig:selfenergyplot2} occurs for a given upper (ultraviolet) limit in the momentum integration $\int_0^\Lambda dk$, and also depends on the effective dimensionless coupling $u = U/(va) \ll 1$.

A straightforward examination of the integral for $\Sigma$,  which we evaluate numerically, reveals that logarithmic contributions are not present. 
Short-range interactions, due to electrostatic contact forces, are typically the most relevant in cold atom systems. Such interactions have been known to affect the spectrum \cite{Dora2013}, and are capable of manipulating and moving the Dirac points, causing opening of a gap or merging the Dirac cones, depending on the interaction sign.

Once the constant term is adjusted to be effectively zero ($\Delta + \Sigma = 0$) due to the above interaction effects, the spectrum is quartic semi-Dirac-like, $\varepsilon({\bf k})=\pm\sqrt{(gk_x^4)^2+(vk_y)^2}$, as shown in Figure \ref{fig:quarticspectrum}. 

Finally, we mention that the cancellation of $\Delta$ is similarly possible under the long range Coulomb interaction, but the momentum-dependent renormalization results in an ordinary semi-Dirac point with quadratically dispersing excitations. The calculation and results are presented in detail in Appendix \ref{AppendixA}. Thus the long range Coulomb interaction is not suitable for stabilizing the quartic semi-Dirac dispersion, and only short range interactions can lead to such a structure in our proposed model. 

\section{Conclusions}
\label{sec:conclusions}

We have shown that it is possible to construct a tight-binding lattice model with an effective low energy quasiparticle spectrum described by exotic quartic semi-Dirac fermions. In order for the quartic spectrum to be stable, one has to introduce electron-electron interactions
to the band structure, which in itself accounts for up to fourth nearest neighbor hopping and anisotropic hopping integrals. We find that short range  interactions are necessary to stabilize the quartic semi-Dirac spectrum, whereas long range Coulomb interaction favor a simple (quadratic) semi-Dirac spectrum. In our view 
the necessary conditions for the emergence of quartic semi-Dirac fermions are more likely achievable experimentally in a cold atom setting due to the tunability of hopping parameters \cite{tarruell2012creating}, and the predominance of short range interactions \cite{Dora2013}. From a theoretical viewpoint it is significant that a microscopic lattice model for higher order semi-Dirac fermions can be constructed. This model will find its place in the list of lattice models describing various Dirac-like excitations. 

\section*{acknowledgments}
We would like to thank Bruno Uchoa for helpful discussions and input. Financial support provided by the University of Vermont is gratefully acknowledged. 

\appendix
\section{Long range Coulomb interactions}
\label{AppendixA}
We consider the long-range Coulomb interaction $e^2/r$ between electrons which in momentum space reads 
$V(\mathbf{k})=2\pi e^2/k,$ where $k=|{\bf k}|$.
The self energy is now given by Eq.(\ref{eq:SE}) with $U$ replaced  by 
$V({\bf k}-{\bf p})$, plus another term obtained from Eq. (\ref{eq:SE}) by replacing $h_{x}(\mathbf{k}) \rightarrow h_{y}(\mathbf{k})$ and $\sigma_x \rightarrow \sigma_y$. 
 Decomposing the self-energy into constant (zero momentum) and finite momentum contributions, the constant part reads
\begin{equation}
\hat{\Sigma}(\mathbf{p}=0)=\left( \iint\frac{\text{d}^{2}{\bf k}}{(2\pi)^2}\frac{h_{x}(\mathbf{k})}{2\varepsilon({\bf k})}\;V({\bf k})\right)\sigma_x,
\label{eq:SE1}
\end{equation}
and the $\sigma_y$ term vanishes by parity. Similarly to Section \ref{sec:interactions}, the scalar piece of the zero momentum contribution, denoted by $\Sigma(\mathbf{p}=0)$, can negate $\Delta$ in Eq.(\ref{generalham}) as shown in Figure \ref{fig:selfenergyplot2}. Again, this occurs for a given ultraviolet cutoff $\Lambda$ depending on the value of the dimensionless coupling $\alpha=e^2/v\ll1$. Suppression of $\alpha$ can be achieved by screening the Coulomb interaction, for example by using substrates with different dielectric constants \cite{Antonio2009}.

As a mathematical curiosity we wish to note that the self energy contains a large $\ln^2$ contribution in the physically relevant regime $\Lambda v/\Delta \gg 1$:
\begin{equation}
\Sigma(\mathbf{p}=0) \approx  \frac{3 \alpha}{4\pi} \Delta \ln^2(\Lambda v/\Delta) + 
e^2\iint\frac{\text{d}^{2}{\bf k}}{4\pi}\frac{g k_x^4}{\varepsilon({\bf k})} .
\end{equation}
The logarithmic part originates from the term proportional to $\Delta$ in  $h_x(\mathbf{k})$.
Calculations with the above formula and the exact numerical evaluation of the integral in Eq. (\ref{eq:SE1}) produce similar results. Such peculiar $\ln^2$ contributions at the one loop, Hartree-Fock level are a characteristic feature of 
strongly anisotropic  semi-Dirac fermions \cite{kotov2021,elsayed2025}.

To study the finite momentum terms in the self energy, we use the expansion
\begin{equation}
V({\bf k}-{\bf p})=\frac{2\pi e^{2}}{\vert{\bf k}-{\bf p}\vert}=2\pi e^{2}\sum_{l=0}^{\infty}\frac{p^{l}}{k^{l+1}}P_{l}(\cos\gamma),
\label{Coulombexpansion}
\end{equation}
where $P_{l}$ are the Legendre polynomials and $\cos\gamma=\frac{{\bf k}\cdot{\bf p}}{kp}$.
The expansion is valid for $k>p$, and only $l=even(odd)$ terms contribute
to the corrections in the $\sigma_x(\sigma_y)$ channel. The leading order term in $\sigma_y$ is a correction to $v$ that has no bearing on our analysis.  We introduce the intrinsic momentum and energy scales
\begin{equation}
    q=\left(\frac{v}{g}\right)^{1/3},\: \varepsilon_0=\left(\frac{v^4}{g}\right)^{1/3}
    \label{scales}
\end{equation}
\begin{figure}
    \centering
    \includegraphics[width=0.9\linewidth]{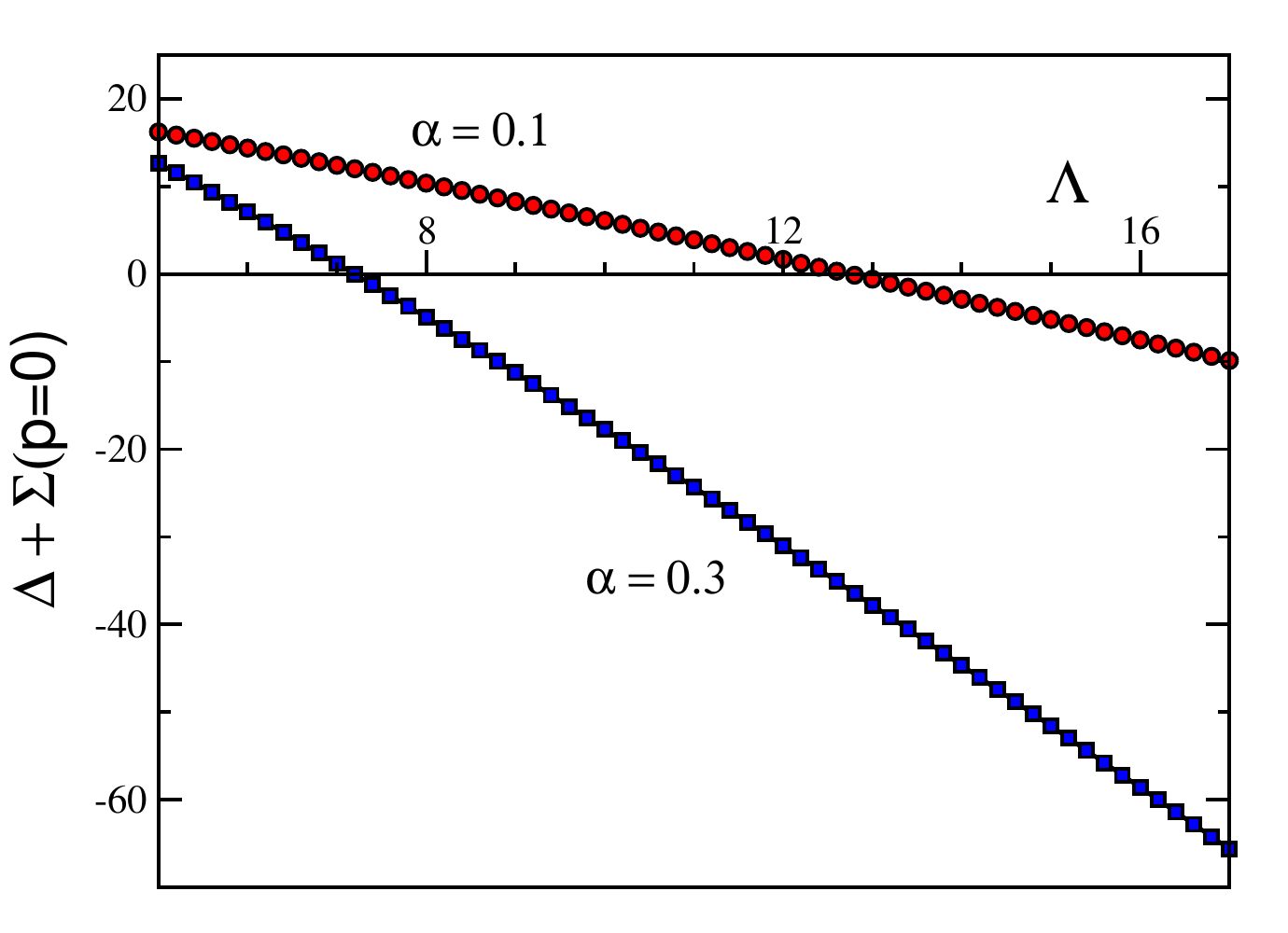}
    \caption{Constant term in the interacting hamiltonian plotted against the ultraviolet momentum
     cutoff $\Lambda$ for different 
    Coulomb interaction strengths $\alpha=0.1, 0.3$. The energies ($y$-axis) are in units of $t_4$ and $\Lambda$ is in units of inverse lattice spacing $a=1$.}
    \label{fig:selfenergyplot1}
\end{figure}
to write the quadratic and quartic contributions to the corrections in $\sigma_x$ as
\begin{align}
    \frac{\alpha}{4\pi}\tilde{p}_x^2\left[\varepsilon_0 I_2(\Delta,\Lambda)-\Delta I_1(\Delta,\Lambda,\lambda)\right]\nonumber\\
    \frac{\alpha}{4\pi}\tilde{p}_x^4\left[\varepsilon_0 I_4(\Delta,\Lambda)-\Delta I_3(\Delta,\Lambda,\lambda)\right],
\end{align}
with the dimensionless integrals
\begin{align}
    I_1&=\int_{\tilde{\lambda}}^{\tilde{\Lambda}}\text{d}\tilde{k}\int_0^{2\pi}\text{d}\theta\frac{P_2(\cos\theta)}{\tilde{k}^2}\frac{1}{\sqrt{[(\tilde{k}\cos\theta)^4-\tilde{\Delta}]^2+(\tilde{k}\sin\theta)^2}}\nonumber\\
    I_2&=\int_{0}^{\tilde{\Lambda}}\text{d}\tilde{k}\int_0^{2\pi}\text{d}\theta\frac{P_2(\cos\theta)}{\tilde{k}^2}\frac{(\tilde{k}\cos\theta)^4}{\sqrt{[(\tilde{k}\cos\theta)^4-\tilde{\Delta}]^2+(\tilde{k}\sin\theta)^2}}\nonumber\\ 
    I_3&=\int_{\tilde{\lambda}}^{\tilde{\Lambda}}\text{d}\tilde{k}\int_0^{2\pi}\text{d}\theta\frac{P_4(\cos\theta)}{\tilde{k}^4}\frac{1}{\sqrt{[(\tilde{k}\cos\theta)^4-\tilde{\Delta}]^2+(\tilde{k}\sin\theta)^2}}\nonumber\\
    I_4&=\int_{0}^{\tilde{\Lambda}}\text{d}\tilde{k}\int_0^{2\pi}\text{d}\theta\frac{P_4(\cos\theta)}{\tilde{k}^4}\frac{(\tilde{k}\cos\theta)^4}{\sqrt{[(\tilde{k}\cos\theta)^4-\tilde{\Delta}]^2+(\tilde{k}\sin\theta)^2}}
\end{align}
where we take the external momentum $\mathbf{p}$ to be along the $x$ direction, $\theta$ is the polar angle of $\mathbf{k}$, and the overscore tilde denotes normalization by the corresponding scale from Eq.(\ref{scales}). Note that while $I_{2,4}$ are finite, $I_{1,3}$, the integrals generated by $\Delta$, diverge in the infrared and hence we introduce a lower momentum cutoff $\lambda$. The range of momenta where quartic behavior is dominant is defined by
\begin{equation}
    \vert p_x\vert\gg q \sqrt{\frac{\vert\varepsilon_0 I_2(\Delta,\Lambda)-\Delta I_1(\Delta,\Lambda,\lambda)\vert}{\vert\varepsilon_0 I_4(\Delta,\Lambda)-\Delta I_3(\Delta,\Lambda,\lambda)\vert}}\equiv p_c
    \label{region}
\end{equation}

\begin{figure}
    \centering
    \includegraphics[width=1\linewidth]{QuarticRegime.pdf}
    \caption{Plot of $p_c=\left(\frac{v}{g}\right)^{1/3}\sqrt{\frac{\vert\varepsilon_0 I_2(\Delta,\Lambda)-\Delta I_1(\Delta,\Lambda,\lambda)\vert}{\vert\varepsilon_0 I_4(\Delta,\Lambda)-\Delta I_3(\Delta,\Lambda,\lambda)\vert}}$ as a function of the lower momentum cutoff $\lambda$ in units of reciprocal lattice spacing. Curves for varying $\Lambda$ are indistinguishable at this scale, and $\Lambda=13$ is taken as the representative value for the ultraviolet cutoff.}
    \label{pc}
\end{figure}
The infrared divergence in $I_{3,4}$ allows us to omit the quartic term from the non-interacting hamiltonian in writing Eq.(\ref{region}), and renders the dependence on $\Lambda$ negligible. The lower cutoff $\lambda$ controls the condition in Eq.(\ref{region}), as is shown in Figure \ref{pc}. Since we use the small $p$ expansion in Eq.(\ref{Coulombexpansion}), the values of $p$ are limited to $p<\lambda$, but $p_c>\lambda$ across the range of $\lambda$ considered. Thus it is evident that the leading behavior is quartic only for momenta outside the range of validity of the integrals $I_{1,3}$, meaning that the Coulomb interaction drives the system to a simple (quadratic) semi-Dirac spectrum. Analysis of higher order terms, for example we have checked the hexic term, shows similar behavior, i.e. eventual dominance of the quadratic term in the infrared. To avoid this issue and obtain a quartic semi-Dirac spectrum, we consider a short range interaction that does not generate a momentum-dependent renormalization, as described in the main text, Section \ref{sec:interactions}.

\nocite{apsrev42Control}
\bibliographystyle{apsrev4-2}
\bibliography{references}
\end{document}